\documentclass{elsart}
\journal{Nuclear Physics A}
\usepackage{epsfig}
\usepackage{amssymb}

\begin{document}

% ------------------------------------------------------------------------- 
% abstract 
% ------------------------------------------------------------------------- 

\begin{frontmatter}

\title{Feasibility study of the observation of the 
neutrino accompanied double beta-decay of $^{76}$Ge to the
$0^{+}_{1}$-excited state of $^{76}$Se using segmented germanium
detectors}
\author[a]{K.~Kr\"oninger},
\author[b]{L. Pandola\corauthref{cor}},
\ead{pandola@lngs.infn.it}
\corauth[cor]{INFN, Laboratori Nazionali del Gran Sasso,
S.S. 17 bis km 18+910, I-67100 L'Aquila, Italy.
Telephone number: +39 0862 437532, Fax number: +39 0862 437570}
\author[c]{V. I. Tretyak}
\address[a]{Max-Planck-Institut f\"ur Physik, M\"unchen, Germany}
\address[b]{INFN, Laboratori Nazionali del Gran Sasso, Assergi (AQ), 
Italy}
\address[c]{Institute for Nuclear Research, Kiev, Ukraine}

\begin{abstract}
Neutrino accompanied double beta-decay of $^{76}$Ge can populate the
ground state and the excited states of $^{76}$Se. While the decay to
the ground state has been observed with a half-life of
$1.74^{+0.18}_{-0.16}\cdot10^{21}$~years, decays to the excited states
have not yet been observed. \\

Nuclear matrix elements depend on details of the nuclear
transitions. A measurement of the half-life of the transition
considered here would help to reduce the uncertainties of the
calculations of the nuclear matrix element for the neutrinoless double
beta-decay of $^{76}$Ge. This parameter relates the half-life of the
process to the effective Majorana neutrino mass. \\

The results of a feasibility study to detect the neutrino accompanied
double beta-decay of $^{76}$Ge to the excited states of $^{76}$Se are
presented in this paper. Segmented germanium detectors were assumed in
this study. Such detectors, enriched in $^{76}$Ge to a level of about
86\%, will be deployed in the \textsc{GERDA} experiment located at the
INFN Gran Sasso National Laboratory, Italy. It is shown that the decay
of $^{76}$Ge to the $1\,122$~keV $0^{+}_{1}$-level of $^{76}$Se can be
observed in {\sc GERDA} provided that the half-life of the process is
in the range favoured by the present calculations which is
$7.5\cdot10^{21}$~y to $3.1\cdot10^{23}$~y.
\end{abstract}

\end{frontmatter}

% ------------------------------------------------------------------------- 
% introduction 
% ------------------------------------------------------------------------- 

\section{Introduction} \label{section:introduction} 

The recent observation of flavor oscillations in experiments with
atmospheric, solar, reactor and accelerator neutrinos has revealed a
non-zero neutrino mass. Being sensitive only to the differences of the
neutrino masses, oscillation experiments yield no information about
the absolute neutrino mass scale or about the nature of the neutrino,
namely if it is a Majorana ($\nu=\overline{\nu}$) or Dirac ($\nu \ne
\overline{\nu}$) particle. Both aspects can be probed by 
neutrinoless double beta-decay (0$\nu\beta\beta$), a process in which
a nucleus of mass $A$ and charge $Z$ transforms into a nucleus with
the same mass and charge $Z+2$ under the simultaneous emission of two
electrons only. This process violates lepton number conservation by
two units and is only possible if the neutrino is a massive Majorana
particle. While $0\nu\beta\beta$-decay is not part of the Standard
Model (SM), it is predicted in many SM extensions, in particular in
grand unified theories (GUTs) and supersymmetric (SUSY) models (see
the latest reviews on double beta decay~\cite{Fae98} and references
therein). Consequently, $0\nu\beta\beta$-decay is sensitive to
different theoretical parameters such as the neutrino mass, lepton
violation constants, right-handed admixtures in the weak currents, the
mass of a right-handed $W_R$ boson and/or other theoretical
parameters, depending on the assumed model. Even the non-observation
of 0$\nu\beta\beta$-decay constrains the parameters of various SM
extensions and narrows the variety of theoretical models. The nuclear
matrix element (NME) which describes the nuclear transition of the
decay is also relevant. If $0\nu\beta\beta$-decay is observed, the
accuracy of the derived theoretical parameters (e.g. the effective
Majorana neutrino mass) also depends on the accuracy of the NME
calculation. The spread between the calculations of NMEs performed
with different methods is still large, and a massive effort is being
devoted to this issue. The current theoretical status and the evidence
of non-vanishing neutrino masses from neutrino oscillation experiments
give a strong motivation for the experimental search for
0$\nu\beta\beta$-decay. \\

Since the first double beta-decay experiment in 1948~\cite{Fir48} the
$0\nu\beta\beta$-decay process is still undetected. The current claim
of discovery of $0\nu\beta\beta$-decay of $^{76}$Ge with a half-life
of $T_{1/2}=1.2^{+3.0}_{-0.5}\cdot10^{25}$~y~\cite{Kla04}
(3~$\sigma$-range) has been widely discussed and still has to be
confirmed by other experiments. Today's best limits on the half-life
are of the order of $10^{24}$~y for $^{130}$Te~\cite{Arn05} and
$^{136}$Xe~\cite{Ber02}, and $10^{25}$~y for
$^{76}$Ge~\cite{Kla01,Aal02}. For about ten other nuclei today's best
limits on the half-life are of the order of $10^{21}-10^{23}$~y (see
compilations~\cite{Tre95,Tre02,Bar06}). \\

The neutrino accompanied double beta-decay ($2\nu\beta\beta$) is a
nuclear transition $(A,~Z)\rightarrow(A,~Z+2)$ which is accompanied by
two electrons and two anti-neutrinos. Such a decay conserves lepton
number and is predicted by the SM. As it is a second-order weak
process it has an extremely long half-life. The measurement of
$2\nu\beta\beta$-decay plays an important role in helping to fix the
nuclear model parameters, in particular the particle-particle strength
parameter, $g_{pp}$, in quasi-particle random phase approximation
(QRPA)-models~\cite{Rod06}. Complementary information can be extracted
from the measurement of the rate of $2\nu\beta\beta$-decay to the
excited states of the daughter nucleus. The probabilities of these
transitions have a different dependence on $g_{pp}$ than the
transition to the ground state~\cite{Suh98,Gri92}. Thus, different
aspects of the nuclear models can be investigated. The observation of
transitions to the excited states can help to constrain the parameter
space in the calculation of NMEs and could hence improve the accuracy
of the calculation of the NME for the 0$\nu\beta\beta$-decay
process. Such an improvement is particularly interesting for $^{76}$Ge
because of the recent claim of discovery of
$0\nu\beta\beta$-decay~\cite{Kla04}. \\

The refinement of experimental methods over the last twenty years
resulted in the observation of $2\nu\beta\beta$-decay. Nowadays,
$2\nu\beta\beta$-decay has been observed in ten isotopes: $^{48}$Ca,
$^{76}$Ge, $^{82}$Se, $^{96}$Zr, $^{100}$Mo, $^{116}$Cd, $^{128}$Te,
$^{130}$Te, $^{150}$Nd, and $^{238}$U, with half-lives in the range of
$10^{18}-10^{21}$~y~\cite{Tre95,Tre02,Bar06}.\\

2$\nu\beta\beta$-decay to excited states is phase-space suppressed. In
only two cases the $2\nu\beta\beta$-decay to the first excited
$0^+_1$-state of the daughter nuclei has been observed,
i.e. $^{100}$Mo with a half-life in the range of
$T_{1/2}=(5.7-9.3)\cdot10^{20}$~y~\cite{100-exc} and $^{150}$Nd with
a half-life of
$T_{1/2}=(1.4^{+0.4}_{-0.2}\,\mathrm{(stat.)}\pm0.3\,\mathrm{(syst.)})\cdot10^{20}$~y~\cite{150-exc}. \\

In this paper a study of the feasibility to detect the
$2\nu\beta\beta$-decay of $^{76}$Ge to the $0^{+}_{1}$-excited state
of $^{76}$Se is presented. The study is performed in the context of
the {\sc GERDA} experiment which will use segmented germanium
detectors in the second phase of the experiment. The segmentation is
the key to the identification of $2\nu\beta\beta$-events.  In
section~\ref{section:segmentation} the {\sc GERDA} experiment and the
design of the segmented germanium detectors are
introduced. Section~\ref{section:2nubbdecay} describes the double-beta
decay of $^{76}$Ge. The signature and event selection are discussed in
section~\ref{section:signatures}. A Monte Carlo simulation developed
in the context of {\sc GERDA} is used to determine the efficiency of
the signal identification as well as the residual background
contributions for different segmentation schemes and is described in
section~\ref{section:simulation}. In section~\ref{section:sensitivity}
the sensitivity of the {\sc GERDA} experiment to this decay process is
presented and discussed. It is shown that the theoretical predictions
of the half-life are within the experimental reach. Conclusions are
drawn in section~\ref{section:conclusions}.

% ------------------------------------------------------------------------- 
% Segmented germanium detectors 
% ------------------------------------------------------------------------- 

\section{{\sc GERDA} and segmented germanium detectors} 
\label{section:segmentation}

The GERmanium Detector Array, {\sc GERDA}~\cite{GERDA}, is a new
experiment which will search for $0\nu\beta\beta$-decay of
$^{76}$Ge. It is currently being installed in the Hall~A of the INFN
Gran Sasso National Laboratory (LNGS), Italy. Its main design feature
is to operate germanium detectors directly immersed in liquid argon
which serves as cooling medium and as a shield against external
$\gamma$-radiation simultaneously. With this setup a background index
of $10^{-3}$~counts/(kg$\cdot$keV$\cdot$y) in the region of the
$Q_{\beta\beta}$-value of $2\,039$~keV is aimed at. \\

Several background reduction techniques have been developed in the
context of {\sc GERDA}. For the first time in double beta-decay
experiments segmented germanium detectors will be deployed. Their
potential for the identification of photons has been
studied~\cite{photonid,siegfried}. The considered segmentation scheme
of the detectors comprises a 3-fold segmentation in the height $z$ and
a 6-fold segmentation in the azimuthal angle~$\phi$. This scheme is
denoted $3_{z}\times6_{\phi}$. Each segment and the core are read out
separately. The detectors will be $n$-type true coaxial germanium
diodes. They are still under design but expected to have a mass of
about 1.6~kg and to be 70~mm high and 75~mm in diameter. This is
slightly smaller than the detectors previously operated by the
Heidelberg-Moscow~\cite{Kla01} and IGEX~\cite{Aal02}
collaborations. The enrichment in $^{76}$Ge is about 86\%. A
non-enriched 18-fold segmented prototype detector has been
successfully operated and
characterized~\cite{siegfried,characterization}. \\

% ------------------------------------------------------------------------- 
% 2nubb-decay  
% ------------------------------------------------------------------------- 

\section{$2\nu\beta\beta$-decay of $^{76}$Ge} 
\label{section:2nubbdecay}

\begin{figure}[ht!]
\begin{center}
\epsfig{figure=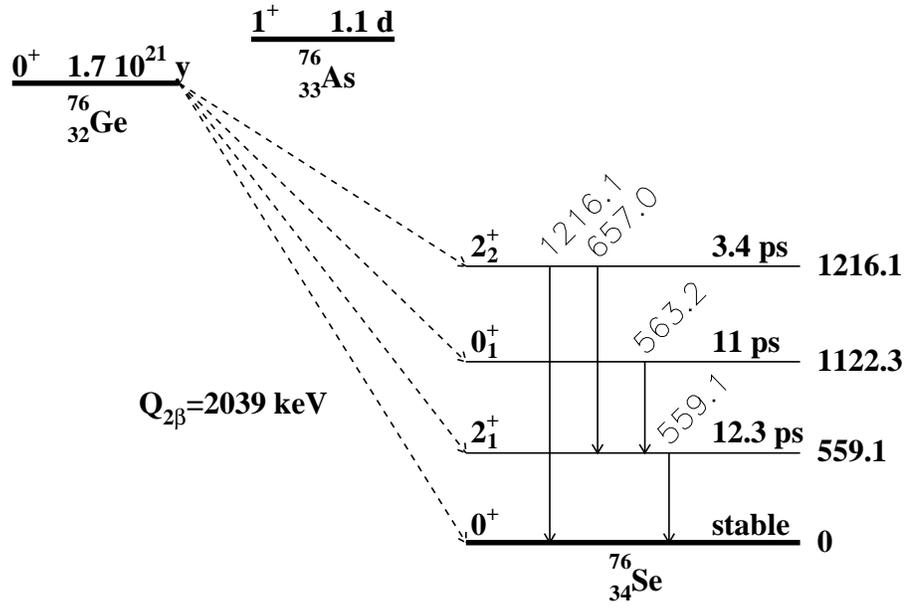,height=8.0cm}
\caption {Lowest energy levels of $^{76}$Se which can be populated 
in the double beta-decay of $^{76}$Ge. The energies of the
excited states and of the de-excitation $\gamma$-rays are given in
keV~\cite{ToI98}.
\label{figure1}}
\end{center}
\end{figure}

The $2\nu\beta\beta$-decay of $^{76}$Ge to the ground state of
$^{76}$Se was observed in several experiments with a measured
half-life in the range of $(0.8-1.8)\cdot10^{21}$~y (see
\cite{Tre02} for references). The most precise determination of the
half-life was achieved in the Heidelberg-Moscow experiment which
yielded a half-life of
$T_{1/2}=1.74^{+0.18}_{-0.16}\cdot10^{21}$~y~\cite{Dor03}.  The
excited states of $^{76}$Se can also be populated. The level scheme of
the double beta-decay of $^{76}$Ge with the lowest energy levels of
$^{76}$Se is shown in Fig.~\ref{figure1}. \\

A summary of the calculated half-lives of the 2$\nu\beta\beta$-decay
of $^{76}$Ge to the excited states of $^{76}$Se is given in
Table~\ref{table1}. The present experimental limits on the half-lives
of these transitions are also listed. The limits cannot test the
calculations so far. Experimentally, the transition to the
$0^{+}_{1}$-level is interesting due to the predicted half-life
between $7.5\cdot10^{21}$~y and $3.1\cdot10^{23}$~y. Transitions to
the $2^+$-levels are suppressed by several orders of magnitude due to
the additional change in spin by two units. \\

\begin{table}[ht!]
\caption{Calculated half-lives and current experimental limits (90\% C.L.) 
for the $2\nu\beta\beta$-decay of $^{76}$Ge to the $2^{+}_{1}$-,
$0^{+}_{1}$- and $2^{+}_{2}$-excited states of $^{76}$Se. Four
different models were used for the calculation, namely the
Hartree-Fock-Bogoliubov model (HFB), the multiple commutator model
(MCM), the quasi-particle random phase approximation (QRPA) and the
shell model (SM).
\label{table1}} 
\begin{center}
\begin{tabular}{lrlllll}
\hline
\multicolumn{2}{c}{Populated}       & \multicolumn{3}{c}{Calculated half-life [y]} & \multicolumn{2}{c}{Experimental limit}        \\
\multicolumn{2}{c}{$^{76}$Se level} & \multicolumn{3}{c}{and model}                & \multicolumn{2}{c}{half-life [y]} \\
\hline
~ & ~ & ~ & ~ & ~ & ~ & ~ \\
2$_1^+$ &  559.1 keV            & $1.2\cdot10^{30}$       & SM   & \cite{Hax84} & $>1.1\cdot10^{21}$ & \cite{Bar95}        \\
~       &  ~                    & $5.8\cdot10^{23}$       & HFB  & \cite{Dhi94} & ~                   & ~                   \\
~       &  ~                    & $5.0\cdot10^{26}$       & QRPA & \cite{Civ94} & ~                   & ~                   \\
~       &  ~                    & $2.4\cdot10^{24}$       & QRPA & \cite{Sto96} & ~                   & ~                   \\
~       &  ~                    & $7.8\cdot10^{25}$       & MCM  & \cite{Aun96} & ~                   & ~                   \\
~       &  ~                    & $1.0\cdot10^{26}$       & MCM  & \cite{Toi97} & ~                   & ~                   \\
~       &  ~                    & $(2.4-4.3)\cdot10^{26}$ & QRPA & \cite{Sch98} & ~                   & ~                   \\
~ & ~ & ~ & ~ & ~ & ~ & ~ \\
0$_1^+$ & $1\,122.3$ keV        & $4.0\cdot10^{22}$       & QRPA & \cite{Civ94} & $>6.2\cdot10^{21}$ & \cite{Vas00}        \\
~       &  ~                    & $7.5\cdot10^{21}$       & MCM  & \cite{Aun96} & ~                   & ~                   \\
~       &  ~                    & $4.5\cdot10^{22}$       & QRPA & \cite{Sto96} & ~                   & ~                   \\
~       &  ~                    & $(1.0-3.1)\cdot10^{23}$ & MCM  & \cite{Toi97} & ~                   & ~                   \\
~ & ~ & ~ & ~ & ~ & ~ & ~ \\
2$_2^+$ & $1\,216.1$ keV        & $1.0\cdot10^{29}$       & QRPA & \cite{Civ94} & $>1.4\cdot10^{21}$ & \cite{Bar95}        \\
~       &  ~                    & $1.3\cdot10^{29}$       & MCM  & \cite{Aun96} & ~                   & ~                   \\
~       &  ~                    & $(0.7-2.2)\cdot10^{28}$ & MCM  & \cite{Toi97} & ~                   & ~                   \\
\hline
\end{tabular}
\end{center}
\end{table}

\clearpage 

% ------------------------------------------------------------------------- 
% Signatures and event selection 
% ------------------------------------------------------------------------- 

\section{Signatures and event selection} 
\label{section:signatures}

\subsection{Signatures}

The $2\nu\beta\beta$-decay of $^{76}$Ge to the $0^+_1$-excited state
of $^{76}$Se at $E^{*} = 1\,122.3$~keV is accompanied by a cascade of
$\gamma$-rays. The final state contains two
anti-neutrinos, two electrons and two $\gamma$-rays. Events
of this kind are referred to as $2\nu\beta\beta-0^{+}_{1}$-events. The
sum energy spectrum of the electrons is continuous with an end-point
at $Q_{\beta\beta} - E^{*} = 917$~keV. The two $\gamma$-rays
$\gamma_{1}$ and $\gamma_{2}$ are monochromatic with energies of
$E_{1}= 559.1$~keV and $E_{2} = 563.2$~keV, respectively.  The
directions of the emitted $\gamma$-rays are correlated. This
correlation can be described as
\begin{equation}
W(\theta) \ = \ \frac{5}{8} \cdot (1 - 3 \cos^{2}\theta + 4\cos^{4}\theta),
\label{correl}
\end{equation}
\noindent 
where $\theta$ is the angle between $\gamma_{1}$ and $\gamma_{2}$ and
$W(\theta)$ is the probability density for $\theta$~\cite{evans}. \\

%\clearpage 

\subsection{Event selection} 
\label{subseection:eventselection}

In the following, three different event selections are introduced for
the identification of $2\nu\beta\beta-0^{+}_{1}$-events. Common to all
three selections is the requirement of a triple-coincidence of three
detector segments. For the 18-fold segmented detectors under study the
two emitted electrons are expected to mostly deposit their energy in
the same segment in which the decay took place. The $\gamma$-rays
emitted in the decay are expected to interact in different segments
due to the longer mean-free path (the interaction length for a 560-keV
$\gamma$-ray in germanium is about 2.5~cm). The event selections are
listed below in increasing restrictiveness:

\begin{itemize}
\item Selection~1: exactly three segments are hit. It is not required 
that the segments belong to the same crystal;
\item Selection~2: exactly three segments are hit. At least one segment 
must show an energy compatible with $\gamma_{1}$ or $\gamma_{2}$;
\item Selection~3: exactly three segments are hit. Two segments must 
show energies compatible with $\gamma_{1}$ and $\gamma_{2}$. In this
case, the event topology is completely known and it is possible to
derive the sum energy spectrum of the two electrons.
\end{itemize}

%\clearpage 

% ------------------------------------------------------------------------- 
% Monte Carlo simulation 
% ------------------------------------------------------------------------- 

\section{Monte Carlo simulation} 
\label{section:simulation}

The simulation is performed using the
\textsc{Geant4}-based~\cite{geant4} \textsc{MaGe} package~\cite{mage}
which is jointly developed and maintained by the {\sc GERDA} and
Majorana Monte Carlo groups. Details on the experimental setup and the
considered physics processes are given in~\cite{photonid}. The
simulated {\sc GERDA} setup includes an array of 21 18-fold segmented
germanium detectors. The number of segments in $\phi$ and $z$ is
varied. \\

The energy resolution of each detector segment and the core are
assumed to be 5~keV full width at half maximum (FWHM). The energy
threshold of each detector segment and the core is assumed to be
50~keV. The assumptions on the energy resolution and the threshold are
conservative with respect to the experimental results achieved with an
18-fold segmented germanium prototype
detector~\cite{siegfried,characterization}. Given the energy
resolution it is not possible to distinguish between $\gamma_{1}$ and
$\gamma_{2}$. The energy range in the event selections~2 and~3 used
for the identification of $\gamma_{1}$- and $\gamma_{2}$-candidates is
$[E_{1}-\textrm{FWHM},E_{2}+\textrm{FWHM}]$. \\

\begin{figure}[ht!]
\center 
\begin{tabular}{cc}
\begin{minipage}[ht!]{0.45\textwidth}
\mbox{\epsfig{file=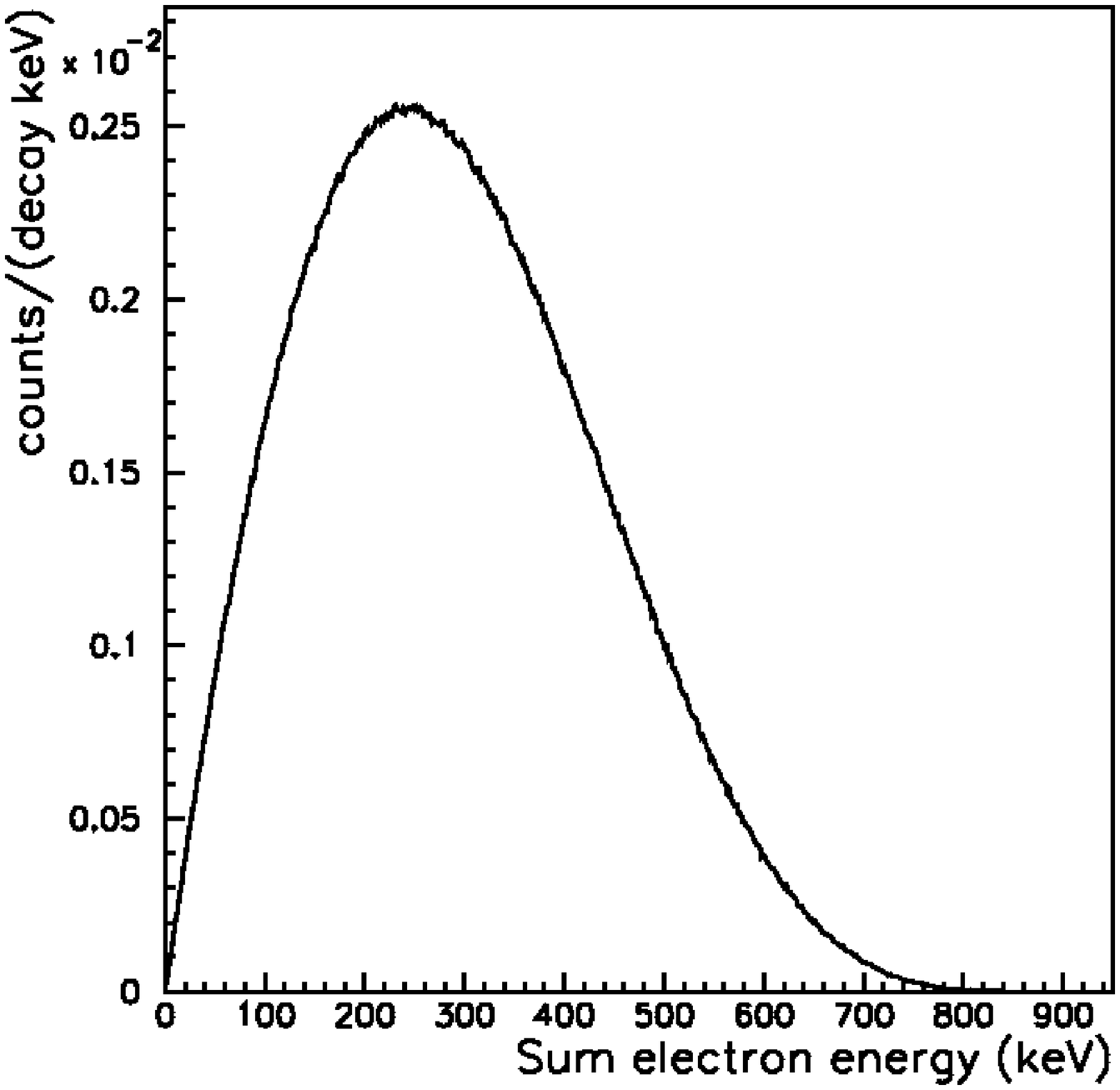,width=\textwidth}}
\end{minipage}
&
\begin{minipage}[ht!]{0.45\textwidth}
\mbox{\epsfig{file=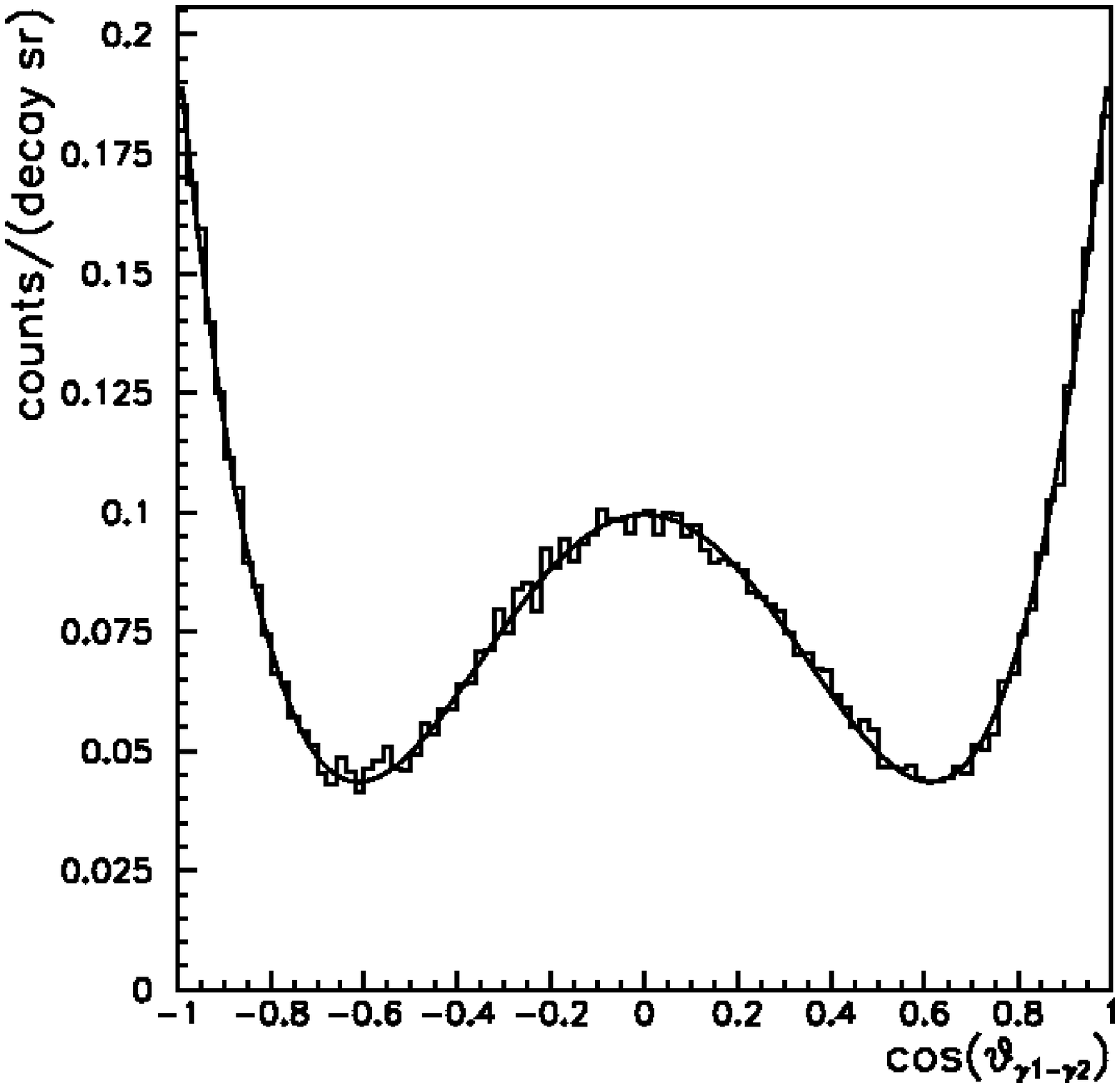,width=\textwidth}}
\end{minipage} \\
\\
\end{tabular}
\caption{Left: Sum energy spectrum of the two electrons emitted 
in the $2\nu \beta\beta$-decay of $^{76}$Ge to the $0^{+}_{1}$-level
of $^{76}$Se. The expected end-point energy is 917~keV.  Right:
Distribution of the angular correlation between the two emitted
photons. Shown are the values derived from the \textsc{decay0} code
(solid histogram) and the theoretical expectation from
Eq.~(\ref{correl}) (dashed curve).
\label{fig:primarykine}}
\end{figure}

For the simulation of the final state of the $2\nu\beta\beta$-decay
the \textsc{decay0} code~\cite{decay0} was used. The code takes the
angular correlation between the two $\gamma$-rays into account
according to Eq.~(\ref{correl}). The sum energy spectrum of the two
electrons is shown in Fig.~\ref{fig:primarykine} (left) as is the
angular distribution between the two $\gamma$-rays (right). The
distributions are derived from the \textsc{decay0} generator. \\

The $2\nu\beta\beta$-events are uniformly distributed inside the
crystals. For each segmentation scheme $10^{5}$ signal events were
generated and the decay products were propagated through the
geometry. Background sources were also simulated. \\

\subsection{Signal detection efficiency}
\begin{figure}[ht!]
\center 
\begin{tabular}{cc}
\begin{minipage}[ht!]{0.45\textwidth}
\mbox{\epsfig{file=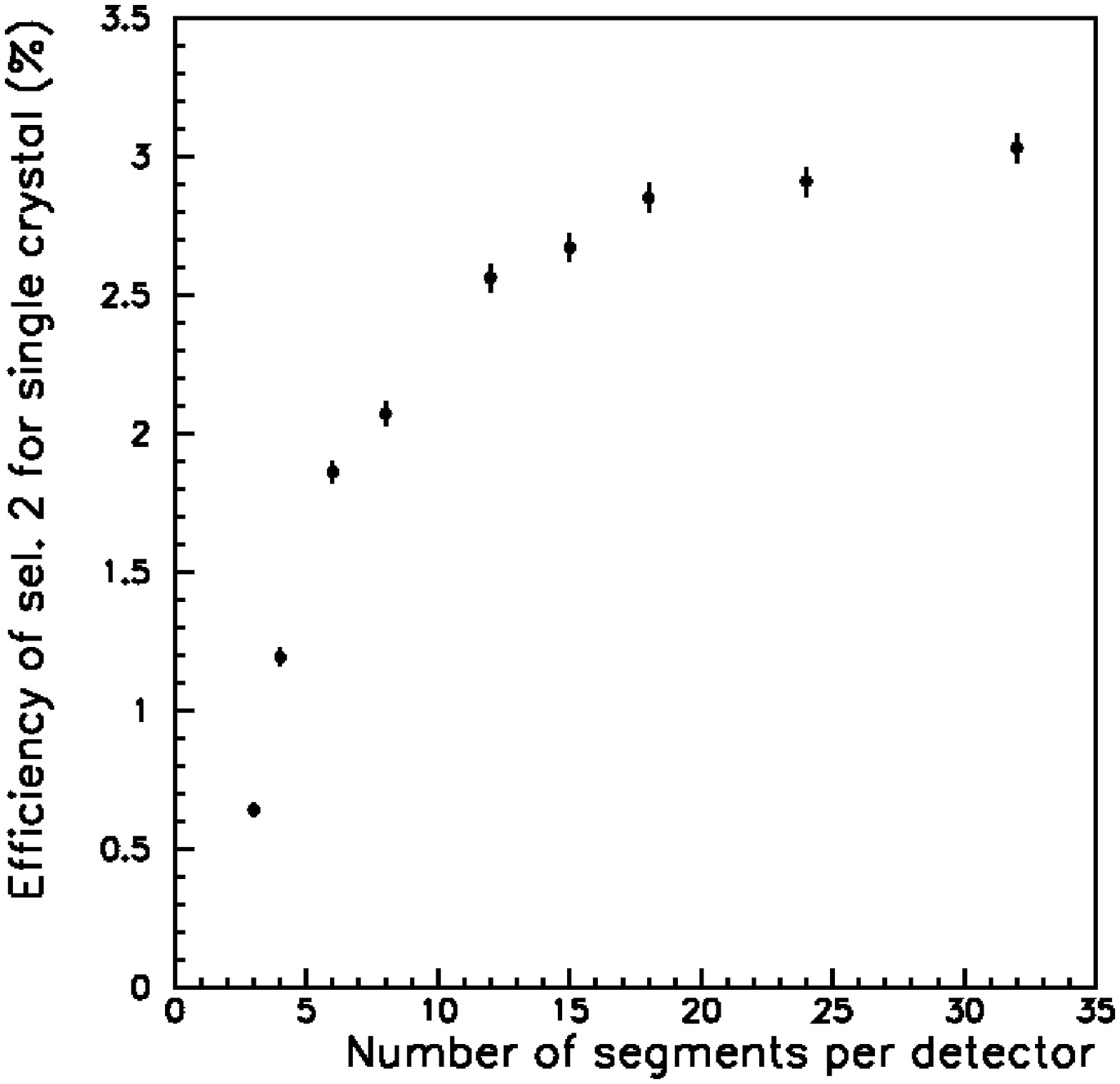,width=\textwidth}}
\end{minipage}
&
\begin{minipage}[ht!]{0.45\textwidth}
\mbox{\epsfig{file=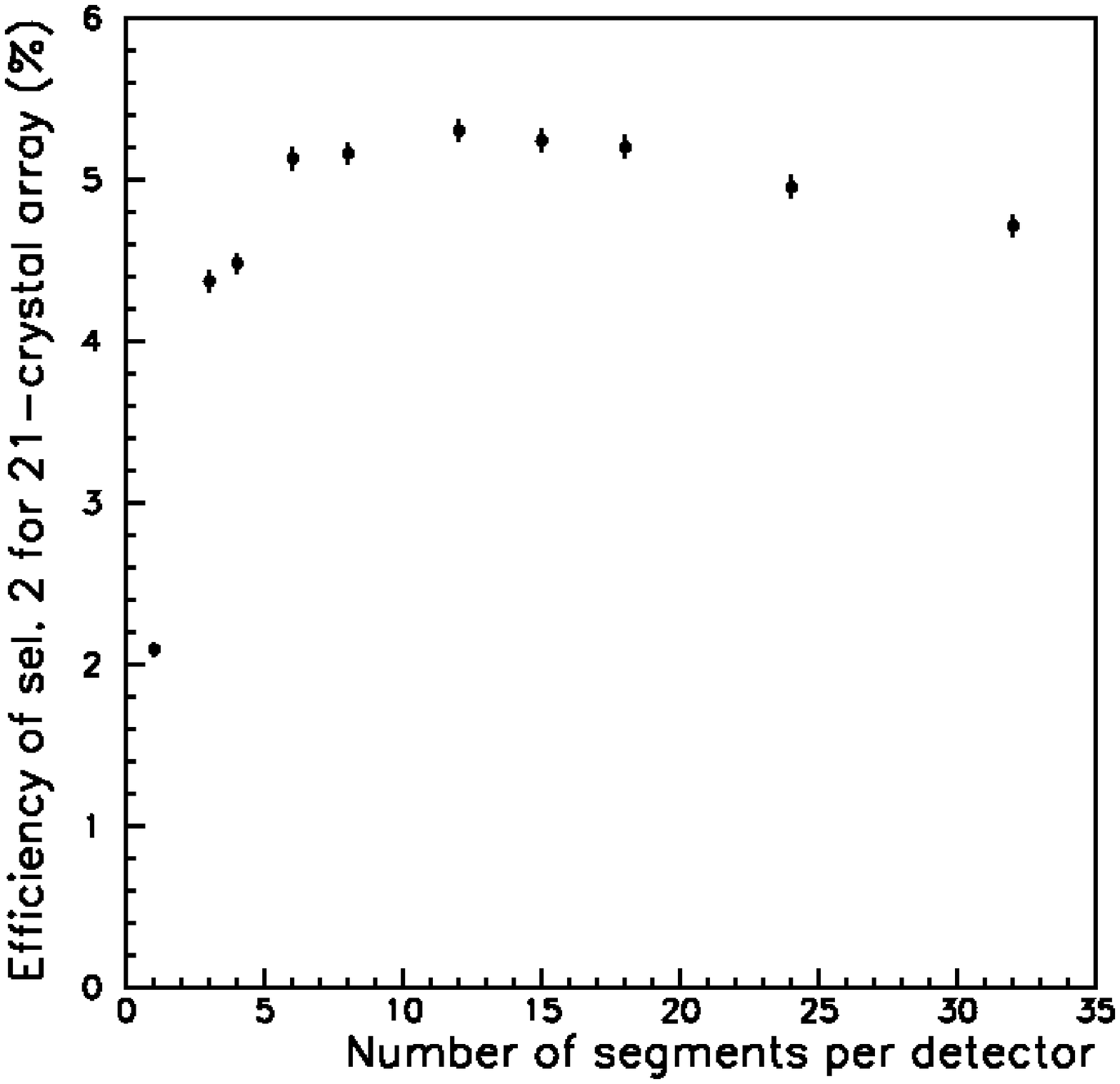,width=\textwidth}}
\end{minipage} \\
\\
\end{tabular}
\caption{Signal detection efficiency for a single, segmented detector 
(left) and for an array of 21 segmented detectors (right) as a
function of the number of segments for event selection~2.
\label{fig:effi}}
\end{figure}

The detection efficiency is defined as the fraction of events which
pass the event selection. The detection efficiency for the
$2\nu\beta\beta-0^+_1$-process, the signal detection efficiency,
depends on the segmentation scheme of the detectors and on the
geometry of the detector array (e.g. on the number and positions of
the detectors, their distance-of-closest approach and the intermediate
material).  Table~\ref{effi_single} summarizes the signal detection
efficiency for a single, segmented $n$-type detector with different
segmentation schemes. For the event selection~2 the detection
efficiency is shown in Fig.~\ref{fig:effi} (left) as a function of the
number of segments. The efficiency increases with the number of
segments until a saturation at around 18~segments is reached. \\

Table~\ref{effi_array} shows the signal detection efficiency for the
array of 21 segmented detectors immersed in liquid argon and different
segmentation schemes. The efficiencies are larger than those for a
single detector because segments of different detectors may be
hit. Fig.~\ref{fig:effi} (right) shows the detection efficiency for
this setup as a function of the number of segments per detector for
event selection~2. The $2\nu\beta\beta-0^{+}_{1}$ selection criteria
can also be met in an array of unsegmented detectors by requiring a
three-fold detector coincidence. In this case, the detection
efficiency is a factor of three smaller than for the reference
$3_{z}\times6_{\phi}$ segmentation scheme foreseen for the {\sc GERDA}
detectors. The detection efficiency initially increases with the
number of segments per detector and forms a plateau between 8 and 18
segments. For a larger number of segments the probability that more
than three segments fire is not negligible and the detection
efficiency decreases. Hence, the reference $3_{z}\times6_{\phi}$
segmentation scheme turns out to be well suited for the identification
of $2\nu\beta\beta-0^{+}_{1}$-events.

\begin{table}[tp]
\caption{Signal detection efficiency for a single, segmented detector 
for different segmentation schemes and event selections. Quoted
uncertainties are statistical only.
\label{effi_single}} 
\begin{center}
\begin{tabular}{c c cc c c}
\hline
\multicolumn{3}{c}{Number of segments} & \multicolumn{3}{c}{Efficiency} \\
along $z$ & along $\phi$ & total & Selection~1 [\%] & Selection~2 [\%] & Selection~3 [\%] \\
\hline
3 & 1 & 3 & $4.25\pm 0.07$ & $0.64\pm 0.03$ & $0.060 \pm 0.008$ \\
1 & 3 & 3 & $4.21\pm 0.07$ & $0.69\pm 0.03$ & $0.08 \pm 0.009$ \\
2 & 2 & 4 & $7.07\pm 0.08$ & $1.19\pm 0.03$ & $0.15\pm 0.01$ \\
3 & 2 & 6 & $11.8\pm 0.1$ & $1.86\pm 0.04$ & $0.21\pm 0.01$ \\
2 & 4 & 8 & $12.6\pm 0.1$ & $2.07\pm 0.05$ & $0.22\pm 0.01$ \\
3 & 4 & 12 & $16.2\pm 0.1 $ & $2.56\pm 0.05$ & $0.32\pm 0.02$ \\
3 & 5 & 15 & $17.5\pm 0.1$ & $2.67\pm 0.05$ & $0.29\pm 0.02$ \\
3 & 6 & 18 & $18.8\pm 0.1$ & $2.85\pm 0.05$ & $0.32\pm0.02$ \\
3 & 8 & 24 & $20.1\pm 0.1$ & $2.91\pm 0.05$ & $0.31\pm 0.02$ \\
4 & 8 & 32 & $21.4\pm 0.1$ & $3.03\pm 0.06$ & $0.34\pm 0.02$ \\
\hline
\end{tabular}
\end{center}
\end{table}
\begin{table}[tp]
\caption{Signal detection efficiency for an array of 21 segmented 
detectors for different segmentation schemes and events
selections. Quoted uncertainties are statistical only. }
\label{effi_array}
\begin{center}
\begin{tabular}{c c cc c c}
\hline
\multicolumn{3}{c}{Number of segments} & \multicolumn{3}{c}{Efficiency} \\
along $z$ & along $\phi$ & total & Selection~1 [\%] & Selection~2 [\%] & Selection~3 [\%] \\
\hline
1 & 1 & 1 & $10.4\pm 0.1$ & $2.09\pm 0.05$ & $0.23\pm 0.01$ \\
1 & 3 & 3 & $22.0\pm 0.1$ & $4.37\pm 0.07$ & $0.56\pm 0.02$ \\
2 & 2 & 4 & $23.3\pm 0.2$ & $4.48\pm 0.07$ & $0.56\pm 0.02$ \\
3 & 2 & 6 & $25.7\pm 0.2$ & $5.13\pm 0.07$ & $0.66\pm 0.03$ \\
2 & 4 & 8 & $26.8\pm 0.2$ & $5.16\pm 0.07$ & $0.64\pm 0.03$ \\
3 & 4 & 12 & $27.6\pm 0.2$ & $5.30\pm 0.07$ & $0.65\pm 0.03$ \\
3 & 5 & 15 & $27.9\pm 0.2$ & $5.24\pm 0.07$ & $0.63\pm 0.03$ \\
3 & 6 & 18 & $28.1\pm 0.2$ & $5.20\pm 0.07 $ & $0.61\pm 0.02$ \\
3 & 8 & 24 & $28.1\pm 0.2$ & $4.95\pm 0.07$ & $0.55\pm 0.02$ \\
4 & 8 & 32 & $27.6\pm 0.2$ & $4.71\pm 0.07$ & $0.53\pm 0.02$ \\
\hline
\end{tabular}
\end{center}
\end{table} 

\clearpage 

\subsection{Background}

Background to the $2\nu\beta\beta-0^{+}_{1}$-process is produced by
the decay of radioactive isotopes inside or in the vicinity of the
detectors. These decays were previously simulated in context of the
$0\nu\beta\beta$-process using the same simulation code. Considered
here are the decays of cosmogenically produced $^{60}$Co and $^{68}$Ge
in the crystals as well as the decays of the radioactive isotopes
$^{238}$U, $^{232}$Th and $^{60}$Co in the suspension and the cabling.
The Monte Carlo data sets are those used in~\cite{photonid} assuming
detectors with an 18-fold segmentation.  Additional data sets were
produced for $^{40}$K and $^{137}$Cs in the corresponding parts close
to the detectors. The unavoidable background due to the
$2\nu\beta\beta$-process to the ground state of $^{76}$Se is also
taken into account. The selection criteria 1--3 were applied to these
data sets to estimate the background contribution. \\

The choice of materials for {\sc GERDA} is not yet final and material
screening results are still pending. The assumed activities are listed
in Table~\ref{table:activities}. Also listed are the activities of
cosmogenically procuded $^{60}$Co and $^{68}$Ge. The total rate of
$2\nu\beta\beta$-decays to the ground state of $^{76}$Se is calculated
to be less than $3\,500$~events/(kg$\cdot$y) assuming a half-life of
$T_{1/2}=1.74\cdot10^{21}$~y. \\

Each detector is assumed to have a mass of 1.6~kg. A detector holder
consists of 31~g copper and 7~g Teflon. The cables for one detector
are assumed to be made out of 1.3~g copper and 0.8~g Kapton.

\begin{table}[tp]
\caption{Assumed activities for the materials close to the detectors in 
units of mBq/kg. Note that screening results for materials to be used
in {\sc GERDA} are still pending. The activity of cosmogenically
produced $^{68}$Ge and $^{60}$Co are 90~events/(kg$\cdot$y) and
5~events/(kg$\cdot$y), respectively. The rate of
$2\nu\beta\beta$-decay events is calculated to be less than
$3\,500$~events/(kg$\cdot$y).
\label{table:activities}} 
\begin{center}
\begin{tabular}{lcccc}
\hline
Isotope              & A(Copper)	& A(Teflon)    & A(Kapton)       & A(Germanium) \\ 
\hline
$^{238}$U             & 0.016  & \phantom{0}0.160 & \phantom{00}9.0 & - \\ 
$^{232}$Th            & 0.012  & \phantom{0}0.160 & \phantom{00}4.0 & - \\ 
$^{137}$Cs            & -      & \phantom{0}0.070 & \phantom{00}3.0 & - \\ 
$^{68}$Ge             & -      & -                & -               & $2.9\cdot10^{-3}$ (see caption) \\
$^{60}$Co             & 0.010  & \phantom{0}0.000 & \phantom{00}2.0 & $1.6\cdot10^{-4}$ (see caption) \\ 
$^{40}$K              & 0.088  & 15.000           & 130.0           & - \\ 
$2\nu\beta\beta$ (g.s.) & -     & -                & -               & 0.111 (see caption) \\ 
\hline
\end{tabular}
\end{center}
\end{table} 

$^{60}$Co events have the highest probability to fake the process
under study because two photons are emitted. As a conservative
estimate the detection efficiencies for events from $^{238}$U,
$^{232}$Th, $^{137}$Cs and $^{60}$Co are set to those of
$^{60}$Co. The efficiency for $^{40}$K is that of events from the
cables. The detection efficiencies for all sources of background are
summarized in Table~\ref{table:background_efficiencies} for the three
event selection criteria. To be conservative the resulting number of
background events are multiplied with a factor 1.5. This yields
background levels of 170.2~events/(kg$\cdot$y),
2.7~events/(kg$\cdot$y) and 0.2~events/(kg$\cdot$y) for the three
event selections, respectively.

\begin{table}[tp]
\caption{Detection efficiencies for the main background contributions obtained 
from the Monte Carlo simulation. The effificiencies for $^{238}$U,
$^{232}$Th, $^{137}$Cs and $^{60}$Co are set to those of
$^{60}$Co. The efficiency for $^{40}$K is that of events from the
cables.
\label{table:background_efficiencies}} 
\begin{center}
\begin{tabular}{lccc}
\hline
Source                & Selection~1 [\%] & Selection~2 [\%]  & Selection~3 [\%] \\ 
\hline 
Cosmogenic $^{68}$Ge  & 27.5             & 0.30              & $0.3\cdot10^{-3}$ \\ 
Cosmogenic $^{60}$Co  & 23.9             & 0.44              & $7.0\cdot10^{-3}$  \\ 
Radioactive $^{60}$Co & 16.1             & 0.32              & $3.0\cdot10^{-3}$ \\  
Radioactive $^{40}$K  & \phantom{0}0.6   & $7.0\cdot10^{-3}$ & $3.0\cdot10^{-3}$ \\ 
$2\nu\beta\beta$ (g.s.) & \phantom{0}0.1   & $2.0\cdot10^{-3}$ & $0.3\cdot10^{-3}$ \\ 
\hline 
\end{tabular}
\end{center}
\end{table} 

% ------------------------------------------------------------------------- 
% Sensitivity
% ------------------------------------------------------------------------- 

\section{Sensitivity} 
\label{section:sensitivity}

The sensitivity is estimated using a statistical analysis method
applied to Monte Carlo data from a simulation of the nominal {\sc
GERDA} setup equivalent to an exposure of 100~kg$\cdot$years.
\subsection{Statistical analysis} 
A Bayesian analysis is used (1) to judge whether the signal process
contributes to the number of observed events and (2) to set a limit on
the signal contribution in case the requirements for a discovery are
not met. The analysis is adopted from that developed
in~\cite{Caldwell:2006yj} for $0\nu\beta\beta$-decay. Here, no
spectral information is used, i.e. only the number of events is used
to evaluate a possible signal contribution. In this formalism the
prior probability for the signal contribution is chosen to be flat in
the number of events up to the current experimental limit which
corresponds to a half-life of $T_{1/2}>6.2\cdot10^{21}$~y, and zero
otherwise. The background is assumed to be known up to Poissonian
fluctuations~\footnote{Residual background for selections~2 and~3 can
be precisely evaluated by slightly shifting the selection window in
energy.}. The prior probabilities for the hypotheses whether or not
the signal process contributes to the data, $H_{2}$ and $H_{1}$,
respectively, are assumed to be equal. The discovery criterion is
defined as in~\cite{Caldwell:2006yj}, namely as $p(H_{1})<10^{-4}$,
corresponding to approximately 3.9~$\sigma$ evidence.

Ensembles of Monte Carlo data were created for fixed signal and
background contributions. In case no discovery can be claimed the 90\%
probability lower limit on the half-life is calculated. The discovery
potential, defined as the half-life for which 50\% of the ensembles
can claim a discovery, is calculated for different exposures.

\subsection{Results}
Table~\ref{table:sensitivity} summarizes the discovery potential and
the expected lower limit on the half-life which can be set in case no
discovery can be claimed. A total exposure of 100~kg$\cdot$years is
assumed. For event selection~2 the sensitivity for the reference
segmentation scheme and an array of unsegmented detectors are
compared. Event selections~1 and~3 are only applied in the case of the
reference segmentation scheme. The signal efficiencies and expected
background levels are taken from section~\ref{section:simulation}. The
background level for the array of unsegmented detectors is expected to
be smaller than that for the array of 18-fold segmented detectors. To
be conservative, the background level for the unsegmented detectors is
assumed to be the same as for the segmented detectors. For the
reference segmentation scheme event selection~2 yields the best
performance with a discovery potential of $T_{1/2}=1.9\cdot10^{23}$~y
or a 90\% probability lower limit on the half-life of
$T_{1/2}>5.6\cdot10^{23}$~y. These results are compatible with the
range of theoretical predictions ($7.5\cdot10^{21}$~y to
$3.1\cdot10^{23}$~y) and about two orders of magnitude above the
present experimental limit of $T_{1/2}>6.2\cdot10^{21}$~y. For an
array of unsegmented detectors the discovery potential and the
expected lower limit are lower by a factor of about 2.5.
\begin{table}[tbh!]
\caption{Discovery potential and expected 90\% probability lower limit on the 
half-life of the $2\nu\beta\beta$-decay of $^{76}$Ge to the
$0^+_1$-excited state of $^{76}$Se for the three event
selections. Detection efficiencies and background levels are taken
from section~\ref{section:simulation}. The background for the array of
unsegmented detectors is conservatively assumed to be the same as for
the array of segmented detectors. A total exposure of
100~kg$\cdot$years is assumed.} \label{table:sensitivity}
\center
\begin{tabular}{lccccc}
\hline
Event selection & Background level         & $T_{1/2}$ discovery & $T_{1/2}$ lower limit \\ 
                & [counts/(kg$\cdot$year)] & potential [y]                           & (90\% prob.) [y] \\ 
\hline 
Sel.~1 ($3_{z}\times6_{\phi}$)   & 170.2                    & $1.3\cdot10^{23}$ & $3.9\cdot10^{23}$ \\ 
Sel.~2 ($3_{z}\times6_{\phi}$)   & \phantom{00}2.7          & $1.9\cdot10^{23}$ & $5.6\cdot10^{23}$ \\ 
Sel.~2 (no segmentation)         & \phantom{00}2.7          & $0.8\cdot10^{23}$ & $2.2\cdot10^{23}$ \\ 
Sel.~3 ($3_{z}\times6_{\phi}$)   & \phantom{00}0.2          & $0.7\cdot10^{23}$ & $2.2\cdot10^{23}$ \\ 
\hline
\end{tabular}
\end{table}

\begin{figure}[ht!]
\center 
\begin{tabular}{cc}
\begin{minipage}[ht!]{0.45\textwidth}
\mbox{\epsfig{file=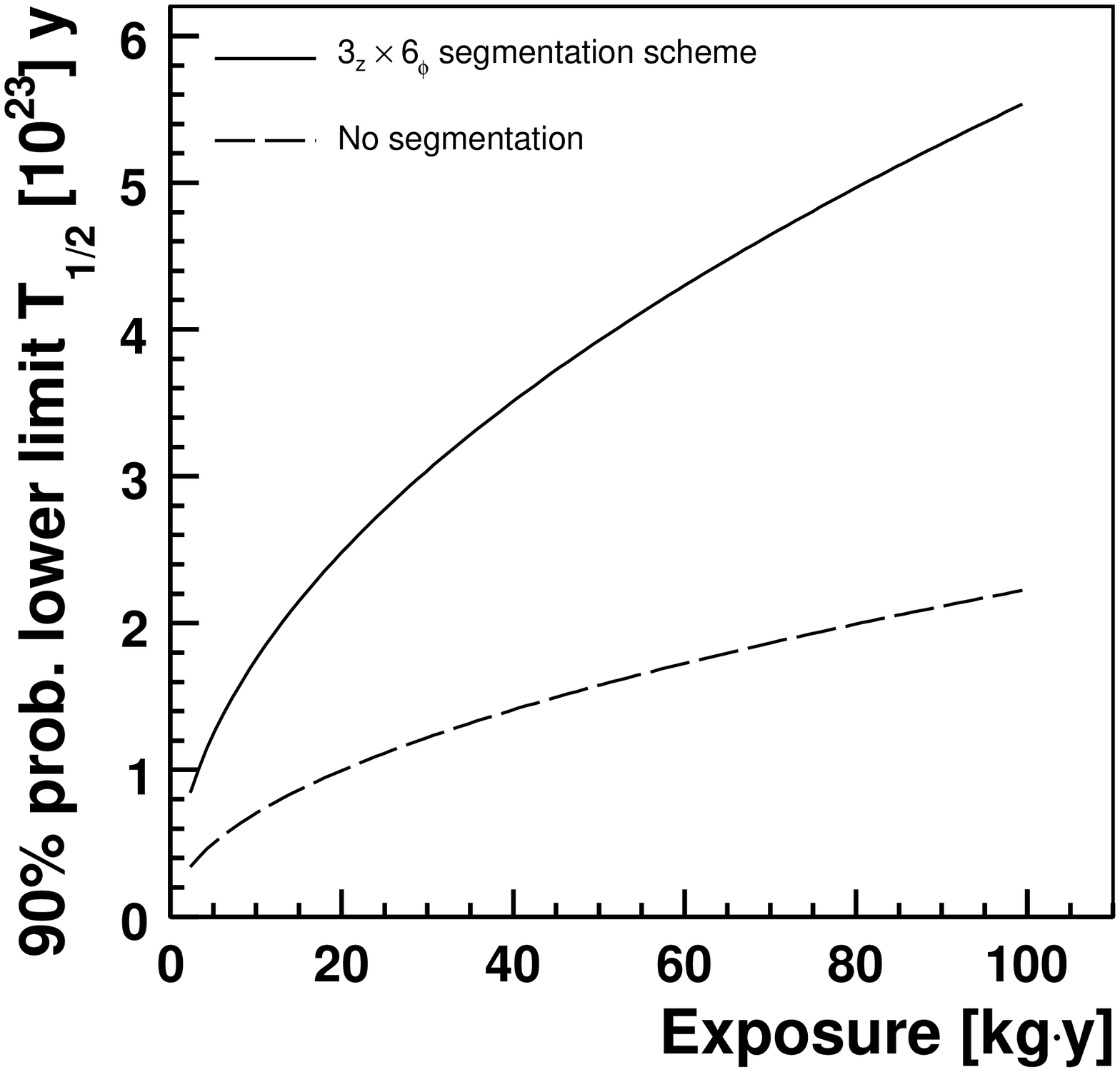,width=\textwidth}}
\end{minipage}
&
\begin{minipage}[ht!]{0.45\textwidth}
\mbox{\epsfig{file=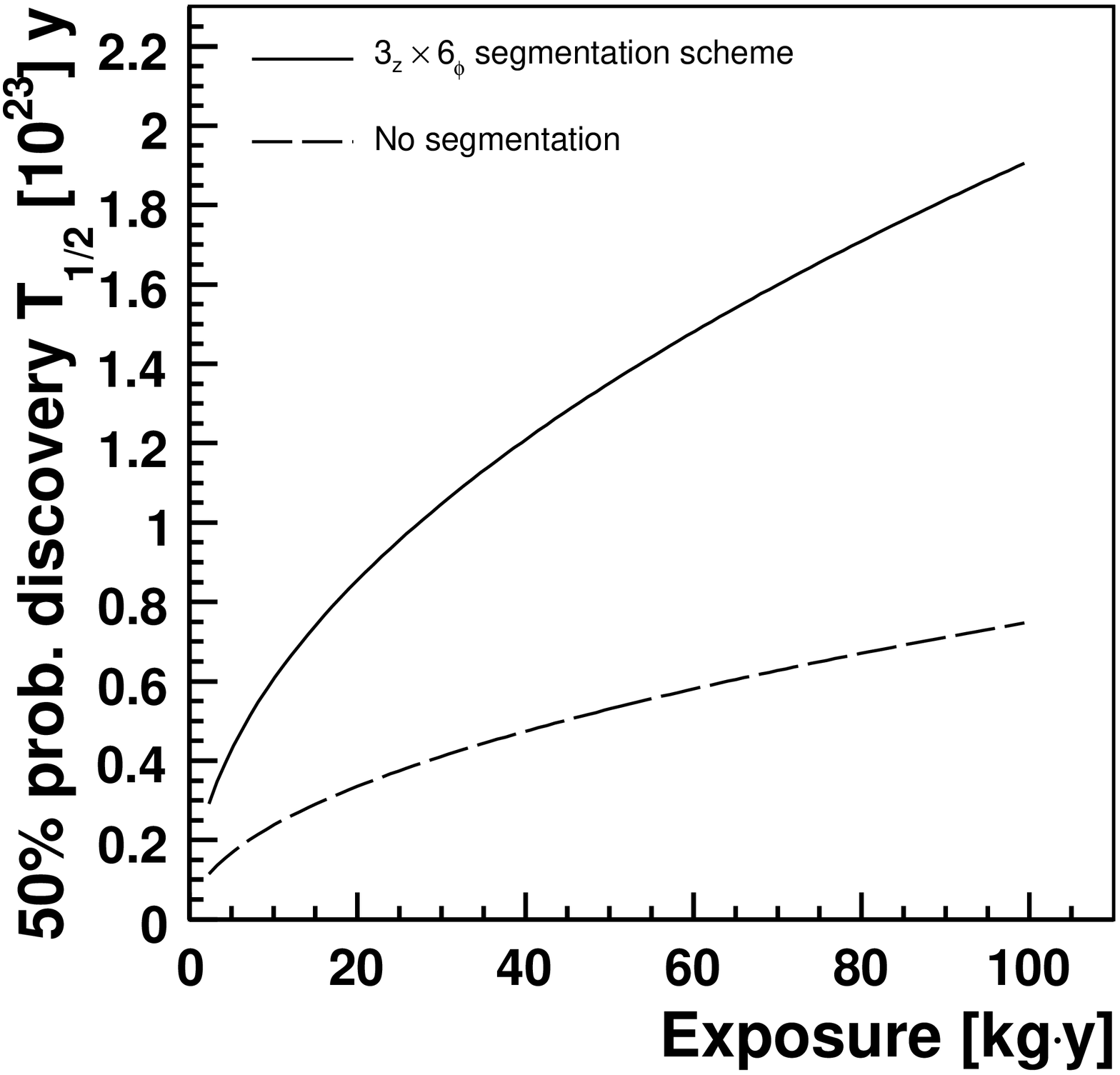,width=\textwidth}}
\end{minipage} \\
\\
\end{tabular}
\caption{Expected lower limit on the half-life (left) and the discovery 
potential (right) for the $2\nu\beta\beta$-decay of $^{76}$Ge to the
$0^+_1$-excited state of $^{76}$Se for the {\sc GERDA} experiment
using event selection~2 as a function of the exposure. The sensitivity
for an array of 18-fold segmented detectors is indicated by the solid
line, the sensitivity for an array of unsegmented detectors is
represented by the dashed line.
\label{fig:sensitivity}}
\end{figure}

Fig.~\ref{fig:sensitivity} shows the expected lower limit on the
half-life (left) and the discovery potential (right) for the double
beta-decay under study for the {\sc GERDA} experiment using event
selection~2 as a function of the exposure. \\

% ------------------------------------------------------------------------- 
% Conclusions 
% ------------------------------------------------------------------------- 

\section{Conclusions} 
\label{section:conclusions}

The study presented indicates the possibility to observe the very
interesting neutrino accompanied double beta-decay of $^{76}$Ge to the
$0^{+}_{1}$-excited state of $^{76}$Se. The 18-fold segmented
germanium detectors to be deployed in the {\sc GERDA} experiment make
it possible to tag single photons in an event and thus identify the
specific decay.  Several event selections and segmentation schemes
were studied. The segmentation scheme considered for the {\sc GERDA}
detectors can improve the sensitivity by a factor of about 2.5
compared to unsegmented detectors yielding a best lower limit on the
half-life of $T_{1/2}>5.6\cdot10^{23}$~y (90\%~prob.). This is two
orders of magnitude above the present experimental limit. A discovery
with 50\% probability or better is expected for half-lives up to
$1.9\cdot10^{23}$~y. This is well within the range favoured by present
calculations which is $7.5\cdot10^{21}$~y to $3.1\cdot10^{23}$~y.

% ------------------------------------------------------------------------- 
% Acknowledgements 
% ------------------------------------------------------------------------- 

\section{Acknowledgments}
\label{section:acknowledgements} 

This work is dedicated to our friends and colleagues M.~Altmann and
N.~Ferrari who prematurely passed away in July, 2006. \\
We express our 
gratitude to Prof. E.~Bellotti for having suggested the possibility to
detect the $2\nu\beta\beta$-decay to excited states using the {\sc
GERDA} segmented detectors, and to Iris Abt and Bernhard
Schwingenheuer for their helpful comments. We would like to thank
R.~Henning, J.~Detwiler and the other colleagues of the Majorana Monte
Carlo group with whom we fruitfully share the development of the
\textsc{MaGe} framework.\\ 
This work has been partially supported by the
\textsc{Ilias} integrating activity (Contract RII3-CT-2004-506222) as
a part of the EU FP6 program.

% ------------------------------------------------------------------------- 
% bibliography 
% ------------------------------------------------------------------------- 

%
%
\end{document}